\documentclass[
 reprint,
 amsmath,amssymb,
prl,superscriptaddress
]{revtex4-2}
\usepackage[version=4]{mhchem}
\usepackage{graphicx,graphics,epsfig}
\usepackage{amsmath, amssymb,amsfonts} 
\usepackage{bm,times,xspace,mhchem}  
\usepackage{mathrsfs}
\usepackage{color}
\usepackage[colorlinks, citecolor=blue, linkcolor=red, citecolor=blue, urlcolor=blue]{hyperref}
\usepackage{dcolumn}
\usepackage{float}
\usepackage{verbatim}
\usepackage{soul}
\usepackage {lineno}

\begin{document}

\preprint{APS/123-QED}

\title{Hybrid Spin and Anomalous Spin-Momentum Locking in Surface Elastic Waves}

\author{Chenwen Yang}
\thanks{
These authors contributed equally: Chenwen Yang, Danmei Zhang, Jinfeng Zhao}
\affiliation{Center for Phononics and Thermal Energy Science, China-EU Joint Lab on Nanophononics, Shanghai Key Laboratory of Special Artificial Microstructure Materials and Technology, School of Physics Science and Engineering, Tongji University, Shanghai 200092, China}

\author{Danmei Zhang}
\thanks{
These authors contributed equally: Chenwen Yang, Danmei Zhang, Jinfeng Zhao}
\affiliation{Center for Phononics and Thermal Energy Science, China-EU Joint Lab on Nanophononics, Shanghai Key Laboratory of Special Artificial Microstructure Materials and Technology, School of Physics Science and Engineering, Tongji University, Shanghai 200092, China}

\author{Jinfeng Zhao}
\thanks{
These authors contributed equally: Chenwen Yang, Danmei Zhang, Jinfeng Zhao}
\affiliation{School of Aerospace Engineering and Applied Mechanics, Tongji University, Shanghai 200092, China}

\author{Wenting Gao}
\affiliation{Center for Phononics and Thermal Energy Science, China-EU Joint Lab on Nanophononics, Shanghai Key Laboratory of Special Artificial Microstructure Materials and Technology, School of Physics Science and Engineering, Tongji University, Shanghai 200092, China}

\author{Weitao Yuan}
\affiliation{School of Mechanics and Aerospace Engineering, Southwest Jiaotong University, Chengdu, Sichuan 610031, China}

 \author{Yang Long}
\affiliation{Center for Phononics and Thermal Energy Science, China-EU Joint Lab on Nanophononics, Shanghai Key Laboratory of Special Artificial Microstructure Materials and Technology, School of Physics Science and Engineering, Tongji University, Shanghai 200092, China}

\author{Yongdong Pan}
\affiliation{School of Aerospace Engineering and Applied Mechanics, Tongji University, Shanghai 200092, China}

\author{Hong Chen}
\affiliation{Center for Phononics and Thermal Energy Science, China-EU Joint Lab on Nanophononics, Shanghai Key Laboratory of Special Artificial Microstructure Materials and Technology, School of Physics Science and Engineering, Tongji University, Shanghai 200092, China}

\author{Franco Nori}
\affiliation{Theoretical Quantum Physics Laboratory, Cluster for Pioneering Research, RIKEN, Wako-shi, Saitama 351-0198, Japan}
\affiliation{Center for Quantum Computing, RIKEN, Wako-shi, Saitama 351-0198, Japan}
\affiliation{Physics Department, University of Michigan, Ann Arbor, MI 48109-1040, USA}

\author{Konstantin Y. Bliokh}
\email{Email: kostiantyn.bliokh@riken.jp}
\affiliation{Theoretical Quantum Physics Laboratory, Cluster for Pioneering Research, RIKEN, Wako-shi, Saitama 351-0198, Japan}
\affiliation{Centre of Excellence ENSEMBLE3 Sp. z o.o., 01-919 Warsaw, Poland}
\affiliation{Donostia International Physics Center (DIPC), Donostia-San Sebasti\'{a}n 20018, Spain}

\author{Zheng Zhong}
\email{Email: zhongk@tongji.edu.cn}
\affiliation{School of Aerospace Engineering and Applied Mechanics, Tongji University, Shanghai 200092, China}

\author{Jie Ren}
\email{Email: Xonics@tongji.edu.cn}
\affiliation{Center for Phononics and Thermal Energy Science, China-EU Joint Lab on Nanophononics, Shanghai Key Laboratory of Special Artificial Microstructure Materials and Technology, School of Physics Science and Engineering, Tongji University, Shanghai 200092, China}

\begin{abstract}
Transverse spin of surface waves is a universal phenomenon which has recently attracted significant attention in optics and acoustics. It appears in gravity water waves, surface plasmon-polaritons, surface acoustic waves, and exhibits remarkable intrinsic spin-momentum locking, which has found useful applications for efficient spin-direction couplers. Here we demonstrate, both theoretically and experimentally, that the transverse spin of surface  elastic (Rayleigh) waves has an anomalous  sign near the surface, opposite to that in the case of electromagnetic, sound, or water surface waves. This anomalous  sign appears due to the hybrid (neither transverse nor longitudinal) nature of elastic surface waves. Furthermore, we show that this sign anomaly can be employed for the selective spin-controlled excitation of symmetric and antisymmetric Lamb modes propagating in opposite directions in an elastic plate. Our results pave the way for spin-controlled manipulation of elastic waves and can be important for a variety of areas, from phononic spin-based devices to seismic waves.
\end{abstract}
\maketitle

 {\it Introduction.---}
Coupling between the spin and orbital degrees of freedom, i.e., between the intrinsic rotation
and external motion of waves or particles, plays an important role in modern physics. It underpins
the field of spintronics~ \cite{Wolf2001,Zutic2004}, topological insulators~\cite{Hasan2010,Qi2011}, and transverse spin-momentum locking
in surface electromagnetic~ \cite{Bliokh2015,Mechelen2016} and acoustic~\cite{Shi2019,Bliokh2019} waves. The later phenomenon describes
the robust link between the direction of propagation of the surface wave and its transverse spin~\cite{Bliokh2015PR,Aiello2015}, i.e., the orthogonal direction of the intrinsic rotation of the wave field or medium particles.
Recently, this effect was employed for highly efficient and reversable spin-to-momentum couplers
using electromagnetic and acoustic surface or guided modes ~\cite{Rodriguez2013,Petersen2014,OConnor2014,Feber2015,Lodahl2017,Shi2019,Xu2020,Long2020,Yuan2021,Han2022}.

\begin{figure}[!t]
\centering
\includegraphics[width=\linewidth]{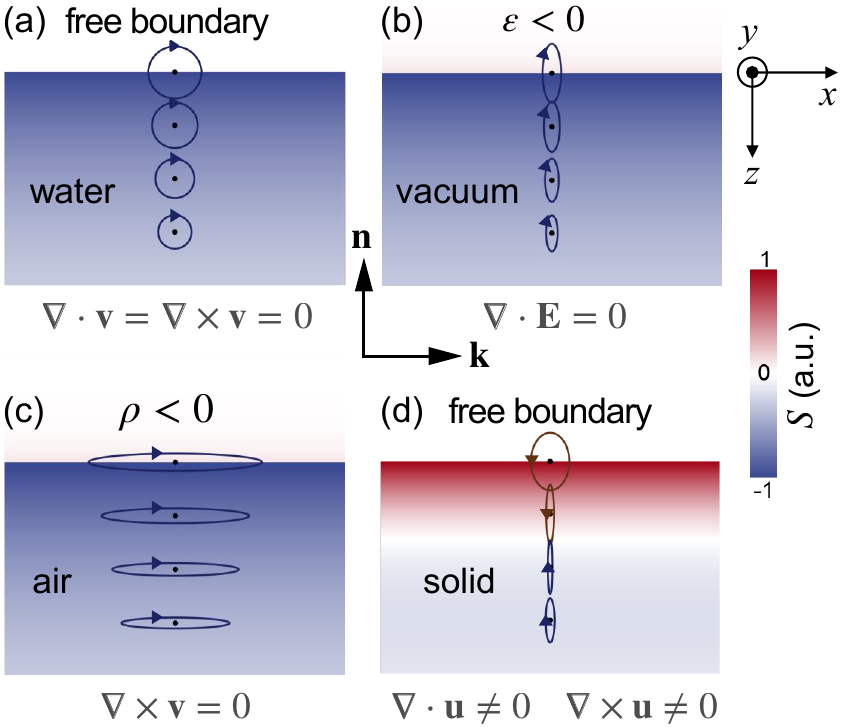}
\caption{ Transverse spin density in surface water (a), electromagnetic (plasmon-polariton) (b), acoustic (c), and elastic Rayleigh (d) waves. Shown are: the $(x,z)$-plane elliptical polarizations of the relevant vector fields for the $x$-propagating surface waves and the corresponding spin densities $S_y \equiv S$. (a) Gravity water waves are described by the divergence-less and curl-less velocity field. (b) Surface plasmon-polaritons at the interface between the vacuum and a negative-permittivity medium (metal) have a transverse (divergence-less) electric field. 
(c) Surface acoustic waves at the interface between air and a negative-density metamaterial have a longitudinal (curl-less) velocity field. 
(d) Elastic Rayleigh waves at the surface of an isotropic solid are described by a {\it hybrid} displacement field which has both longitudinal and transverse contributions, Eq.~(\ref{eq15}).}
\label{fig1}
\end{figure}

The best known example of the transverse spin and spin-momentum locking is the rotation
of water particles in water-surface gravity waves~\cite{LLfluid}, Fig.~\ref{fig1}(a). The normal to the water surface
(or the wavefield intensity gradient) $\bf n$, the wave propagation direction (momentum or wave vector) $\bf k$, and the local water-particle rotation (spin angular momentum~\cite{Jones1973,Longuet1980}) $\bf{s}$ always form a
right-handed triad: ${\bf s} \cdot ({\bf n} \times {\bf k}) > 0$. As a result, the reversal of the wave propagation direction inevitably reverses the transverse spin (the particles' rotation direction) and vice versa. Similar spin-momentum locking occurs in surface electromagnetic (surface plasmon-polaritons) \cite{Maier_book} and acoustic \cite{Ambati2007,Park2011} waves involving locally rotating electric field and medium particles, respectively \cite{Bliokh2015,Mechelen2016,Shi2019,Bliokh2019,Bliokh2015PR,Aiello2015,Rodriguez2013,Petersen2014,OConnor2014,Feber2015,Lodahl2017,Shi2019,Xu2020,Long2020,Yuan2021}. 
{Despite profound differences between the transverse (divergence-less) electromagnetic waves, longitudinal (curl-less) acoustic waves, and water gravity (divergence-less and curl-less) waves, the $(\bf{n},{\bf k},{\bf s})$ triad is always right-handed, as shown in Fig.~\ref{fig1}(a-c).}

It is well known that surface elastic (Rayleigh) waves also involve elliptical orbits of the
medium particles~\cite{Fang_book,Auld_book,Stein_book} and thus possess a transverse spin~\cite{Long2018,Nakane2018,Bliokh2022,Ren2022}. The efficient spin-to-direction coupling using the spin-momentum locking in elastic waves was recently demonstrated
~\cite{Yuan2021}; and the rotational motion of the particles in a surface acoustic wave was detected via a rather sophisticated indirect method~\cite{Sonner2021}. 
 In addition, the ellipticity of the near-surface particle motion was studied in seismic Rayleigh waves \cite{Boore1969,Ferreira2007,Tanimoto2008,Lin2014,Berbellini2016}.

Remarkably, the spin-normal-momentum triad is {\it left-handed} near the surface for Rayleigh waves:
 ${\bf s} \cdot ({\bf n} \times {\bf k}) < 0$, Fig.~\ref{fig1}(d). In comparison with water waves, the opposite rotation of the elastic medium particles is sometimes called ``retrograde'' \cite{Fang_book, Stein_book}. Furthermore, the direction of the particles' rotation, i.e., the spin sign, {\it flips} deep into the elastic medium. 
 
 In this Letter, we explain these spin-sign anomalies in surface elastic waves by their {\it hybrid} nature with both longitudinal (curl-less) and transverse (divergence-less) contributions to the wavefield. 
Such anomalies do not appear in the case of purely longitudinal or transverse waves. 
We experimentally measure these spin anomalies and provide the first, to the best of our knowledge,  measurements of the reversal of the Raleigh-wave spin as a function of the depth.
Moreover, we show that the anomalous spin sign in hybrid waves can provide novel functionalities to setups based on the spin-momentum locking. Namely, we demonstrate the selective excitation of oppositely-propagating symmetric and antisymmetric Lamb modes in an elastic plate by a chiral source.

Our results shed light onto the nature of the anomalous behaviour of the transverse spin
in elastic waves, provide a new tool for elastic spin-orbit coupling manipulations, and can find
applications in condensed-matter systems involving phonon spin \cite{Uchida2011,Weiler2012,Kobayashi2017,Puebla2022}, elastic metamaterials~\cite{Mousavi2015,Cummer2016,Krushynska2023,phononic_skyrmion}, and various seismic-waves studies \cite{Boore1969,Ferreira2007,Tanimoto2008,Lin2014,Berbellini2016,Brule2014,Connell2000,Stein_book}.

{\it Spin sign anomalies in surface Rayleigh waves.---}
The absolute and normalized densities of the spin angular momentum in monochromatic acoustic waves in  nondispersive fluids or solids can be written as~\cite{Long2018,Nakane2018,Shi2019,Bliokh2019,Jones1973,Bliokh2022SA,Ren2022,Bliokh2022}:
\begin{equation}
\begin{aligned}
{\bf S}= \frac{\rho\, \omega}{2}\, {\rm Im}(\bf{u}^* \times \bf{u})\,, \quad
\bf{s}= \frac{{\rm Im}(\bf{u}^* \times \bf{u})}{\left|\bf{u}\right|^2} 
\end{aligned}
    \label{eq.SAM}
\end{equation}
Here $\rho$ is the mass density of the medium, $\omega$ is the wave frequency, $\bf{u}({\bf r})$ is the wave-induced complex displacement amplitude of the medium particles, whereas the real time-dependent displacement is given by ${\rm Re}({\bf u}\, e^{-i \omega t})$. 
Equation (\ref{eq.SAM}) has a universal form, wherein ${\bf S}$ represents the real spin angular momentum of elastic vibration that is related to the mechanical torque, and $|{\bf s}| \leq 1$ describes the ellipticity of the field $\bf u$~\cite{Berry2001}. Both ${\bf S}$ and ${\bf s}$ are directed along the normal to the polarization ellipse of $\bf u$. 
This equation also describes the spin in water surface waves~\cite{Longuet1980,Bliokh2022SA}, while the spin density in electromagnetic waves is expressed similarly with the electric wavefield $\bf E$ instead of the particle velocity $ - i \omega {\bm u}$ and the medium permittivity $\varepsilon$ instead of $\rho$ ~\cite{Bliokh2015PR,Aiello2015}.

Consider a surface wave propagating along the $x$-direction at the $z = 0$ surface and having the form $\propto \exp(ik_x x - \kappa z), k_x > 0, \kappa > 0,$ in the medium $z > 0$. Purely transverse waves (e.g., surface plasmon-polaritons with ${\nabla} \cdot {\bf E} = 0$ \cite{Maier_book}), purely longitudinal waves (e.g., surface acoustic waves with ${\nabla} \times {\bf u} = 0$ \cite{Ambati2007,Park2011}), and simultaneously transverse and longitudinal waves (e.g., gravity water waves with ${\bm \nabla} \cdot {\bf u} = {\nabla} \times {\bf u} = 0$ \cite{LLfluid}) are all characterized by an elliptical polarization of the wavefield in the propagation $(x, z)$ plane and the corresponding transverse spin $S_y \equiv S < 0$ forming the right-handed triad $({\bf s}, {\bf n}, {\bf k})$ \cite{Bliokh2015PR,Aiello2015,Lodahl2017,Shi2019,Bliokh2019,Bliokh2017PRL}, see Fig.~\ref{fig1}(a--c) 
\footnote{
For simplicity, in Figs.~\ref{fig1}(b,c) we highlight the transverse spin density of surface electromagnetic and acoustic waves in free space ($\varepsilon =1$) and air ($\rho >0$), respectively. One can equally consider the wave spin inside negative-index media with $\rho(\omega) < 0$ and $\varepsilon (\omega) < 0$. In this case, the negative parameter in Eq. (1) should be substituted by its dispersion-modified positive version,
$\rho \rightarrow \tilde{\rho} > 0$ or $\varepsilon \rightarrow \tilde{\varepsilon} > 0$, which ensures that the directions of {\bf S} and {\bf s} always coincide~\cite{Bliokh2019, Bliokh2017PRL}.
}. 
Here and hereafter we consider only 2D $(x,z)$ fields and the transverse spin having only the $y$-component, omitting the index $y$ for the sake of brevity.
Surprisingly, the transverse spin of surface elastic Rayleigh waves behaves quite differently, as shown in Fig.~\ref{fig1}(d) \cite{Fang_book,Auld_book,Stein_book}. First, $S > 0$ near the surface, so that $({\bf s}, {\bf n}, {\bf k})$ is a left-handed triad. Second, the spin sign flips at some distance from the surface: $S < 0$ for $z > z_c > 0$. 

\begin{figure*}[!t]
  \centering
  \includegraphics[width=\linewidth]{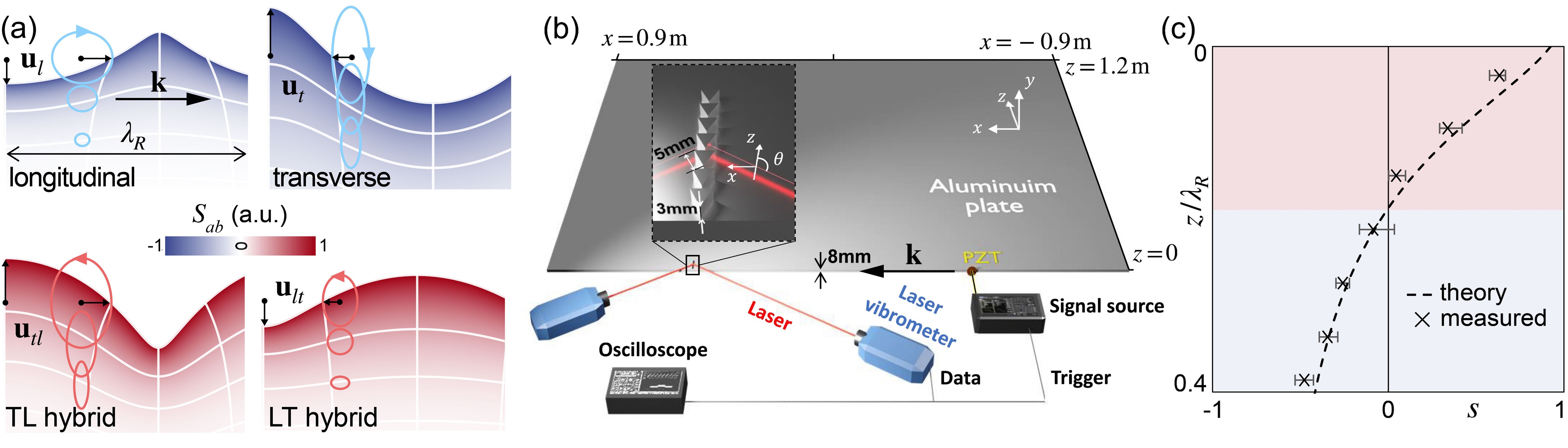}
  \caption{ (a) Rayleigh-wave-induced deformations of an elastic medium caused by the purely longitudinal and transverse displacement field parts ${\bf u}_l$ and ${\bf u}_t$, as well as by the mixed fields ${\bf u}_{tl}=(u_{t\,z}, u_{l\,x})$ and ${\bf u}_{lt}=(u_{l\,z}, u_{t\,x})$ (the latter ones are not real displacement fields but rather visualized contributions to the total spin density). The corresponding field polarizations are shown by ellipses, whereas the spin densities are indicated by the red-blue color scheme. 
(b) The experimental setup for the measurements of the Rayleigh-wave spin (see explanations in the text).
(c) The measured versus calculated normalized spin density $s_y \equiv s$ of the $x$-propagating Rayleigh wave. The Rayleigh wavelength is $\lambda_R\approx 8$ cm at 35 kHz. The background colors highlight the areas of positive and negative spin, cf. Fig.~\ref{fig1}(d).
\label{fig2}}
\end{figure*}

To explain these two anomalies, we note that the Rayleigh wave is formed by a {\it hybrid} of longitudinal (compression) and transverse (shear) oscillations of the medium. Its wavefield can be expressed as a sum of longitudinal and transverse contributions: 
\begin{equation}
{\bf u} = {\bf u}_l + {\bf u}_t\,, \quad 
{\nabla} \times {\bf u}_l = 0\,,~~{\nabla} \cdot {\bf u}_t = 0\,.
\label{eq15}
\end{equation}
These two contributions have different amplitudes and exponential decay rates: ${\bf u}_{l,t} = {\bf A}_{l,t} \exp (ik_x x - \kappa_{l,t} z)$. Substituting Eq.~(\ref{eq15}) into Eq.~(\ref{eq.SAM}), we find that there are purely longitudinal, purely transverse, and `hybrid' contributions to the Rayleigh-wave spin~\cite{Long2018}:
\begin{align}
{\bf s} = \frac{1}{|{\bf u}|^2} {\rm Im}\!\left[({\bf u}_l^* + {\bf u}_t^*)\times ({\bf u}_l + {\bf u}_t)\right] 
=  {\bf s}_{ll} + {\bf s}_{tt} + {\bf s}_{lt} + {\bf s}_{tl} \,.
\label{eq2}
\end{align}
For 2D fields under consideration, $s_{ab} = 2\, {\rm Im}({u}_{a\,z}^* {u}_{b\,x})/|{\bf u}|^2$, i.e., the `pure' contributions $s_{ll}$ and $s_{tt}$ are determined by the elliptical polarizations of the fields ${\bf u}_l$ and ${\bf u}_t$, while the hybrid contributions $s_{lt}$ and $s_{tl}$ are determined by the elliptical polarizations of the `mixed' fields ${\bf u}_{lt}=(u_{l\,z}, u_{t\,x})$ and ${\bf u}_{tl}=(u_{t\,z}, u_{l\,x})$,  respectively. The polarizations of these pure and mixed fields are shown in Fig.~\ref{fig2}(a). Remarkably, the pure contributions to the spin are always negative: $s_{ll} < 0$, ${s}_{tt} < 0$, while the hybrid contributions are positive: ${s}_{lt} > 0$, ${s}_{tl} > 0$ (see Supplementary Materials~\footnote{See Supplemental Material, which includes Refs.~\cite{Maier_book,Bliokh2015,Ambati2007,Park2011,Shi2019,Bliokh2019,Graff_book,Rose_book}\label{fn}}). Thus, it is the balance of these contributions that determines the sign of the Rayleigh-wave spin. Near the surface, the hybrid contributions always prevail and the spin becomes positive. However, the longitudinal, transverse, and hybrid contributions contain different $z$-dependences: $\propto \exp(-2\kappa_l z)$, $\propto \exp(-2\kappa_t z)$, and $\propto \exp[-(\kappa_l+\kappa_t) z]$, respectively. Since $\kappa_t < \kappa_l$, the hybrid contributions decay faster than the transverse one, and the sum of pure contributions start to prevail after some $z = z_c$ 
{($z_c \simeq 0.2 \lambda_R$, where $\lambda_R = 2\pi/k_x$ is the wavelength of the Rayleigh wave)}. Thus, the anomalous spin sign at $z=0$ can be attributed to the existence of hybrid contributions, while the reversal of the spin direction at $z = z_c$ is due to the faster decay of the hybrid contributions compared to the pure ones.

Although 
the near-surface spin/ellipticity in surface acoustic \cite{Yuan2021,Sonner2021,Xu2020} and seismic \cite{Boore1969,Ferreira2007,Tanimoto2008,Lin2014,Berbellini2016} waves has been measured, its $z$-dependence, sign reversal, and pure/hybrid contributions have never been observed experimentally. We performed experimental measurements of the
$(x, z)$-plane vibrations in a Rayleigh wave with frequency $f= \omega/2 \pi=35\,$kHz propagating along the surface of an aluminum plate. The experimental setup is shown in Fig.~\ref{fig2}(b). 
The $x$-propagating Rayleigh wave was excited by a PZT piezoelectric ring attached to the edge of the plate, $z=0$, where the electric signal was generated by the function generator (RIGOL DG1032z)  and then amplified by the power amplifier (Aigtek ATA-2022H). The plate thickness along the $z$-axis is $1.2\,$m, which can be  considered as infinite compared with the skin depth of the Rayleigh wave at 35 kHz.

To observe the spin-sign flip, we measured the 2D displacement field ${\bf u}$ at different $z$-points. In order to do this in an opaque material (aluminium) in the ultrasonic frequency range (which is close to the application of surface-acoustic-wave devices), we used vibrations of seven triangular prisms attached to the original plate at $z = (0.3, 0.8, 1.26, 1.7, 2.2, 2.7, 3.1)\,$cm, see the inset in Fig.~\ref{fig2}(b). By measuring the normal displacements of the two sides of each prism using two lasers Doppler vibrometers (Polytec OFV 2570) we extracted the 2D displacement vector. Namely, for the two lasers directed at angles $\theta_1$ and $\theta_2$ with respect to the $z$-axis and measuring the vibration signal $u_1$ and $u_2$, the resulting displacement vector has components $u_x = u_1 \sin \theta_1 + u_2 \sin \theta_2$ and $u_z = u_1 \cos \theta_1 + u_2 \cos \theta_2$.

Figure~\ref{fig2}(c) shows the results of these measurements of the $z$-dependence of the normalized spin density $s = 2\,{\rm Im}(u_z^* u_x)/\!\left( \left|u_x\right|^2+\left|u_z\right|^2 \right)$. It clearly exhibits the spin-sign anomalies in perfect agreement with the theoretical calculations. We have also retrieved the separate  contributions $s_{ab}$, which explain the origin of the nontrivial behavior of the Rayleigh-wave spin, from the measurements of the aluminium-plate strain. These results are shown in the Supplementary Materials~\footnotemark[2].

\begin{figure}[!t]
  \centering
  \includegraphics[width=\linewidth]{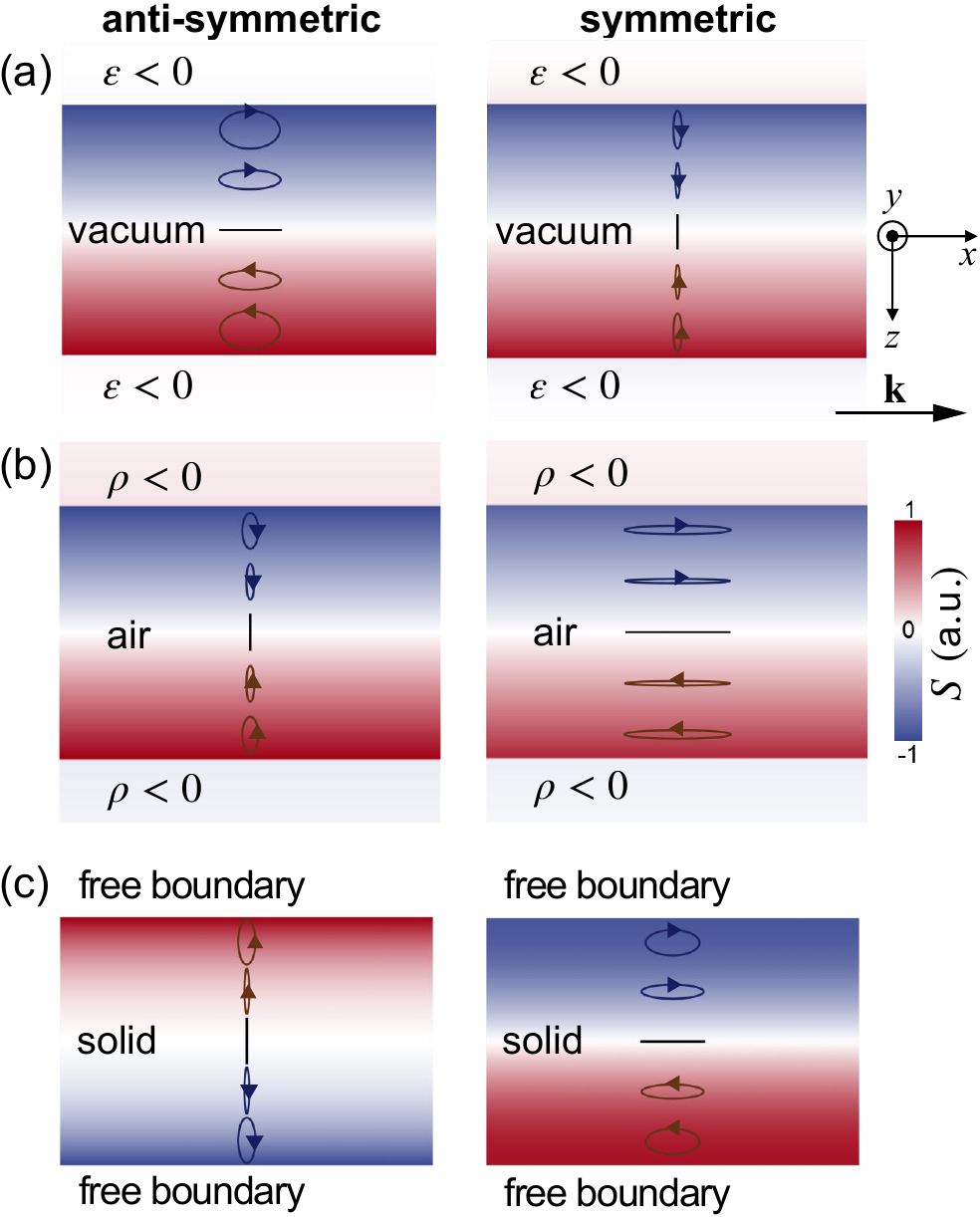}
 \caption{ Transverse spin density in the  lowest-order anti-symmetric (A0) and symmetric (S0) modes in electromagnetic (a), acoustic (b), and elastic (c) thin-plate/slit systems, cf. Figs.~\ref{fig1}(b)--(d). The A0 elastic Lamb mode has a reversed spin sign compared to its electromagnetic and acoustic counterparts.}
\label{fig3}
\end{figure}

\begin{figure}[!t]
  \centering
  \includegraphics[width=0.93\linewidth]{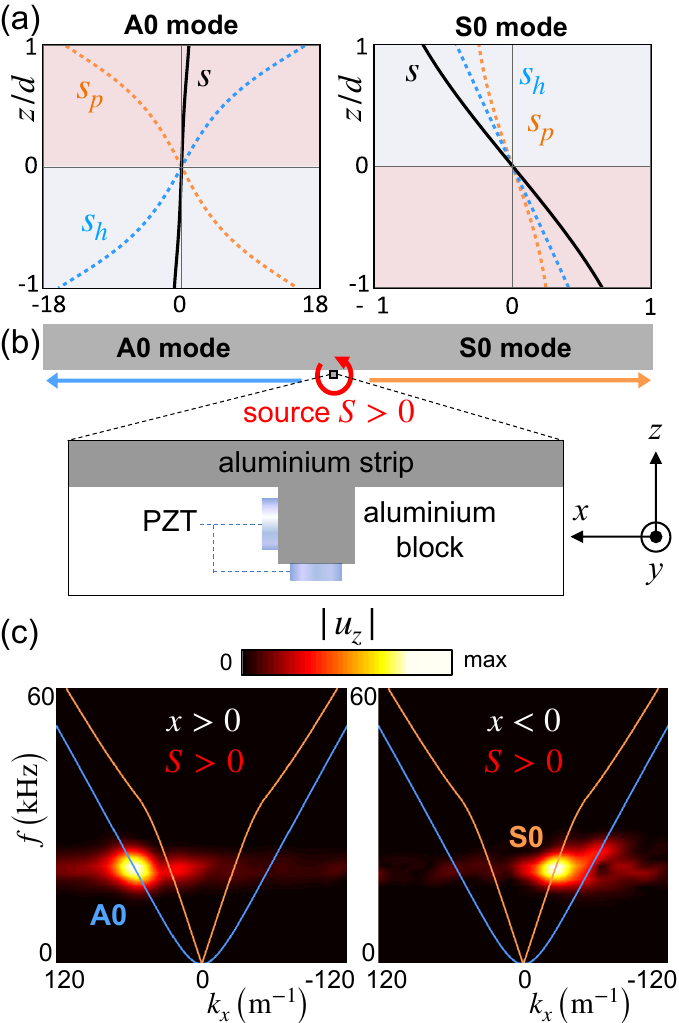}
  \caption{(a) The pure, $s_{p} = s_{ll} + s_{tt}$, and hybrid, $s_{h} = s_{lt} + s_{tl}$, contributions to the normalized spin $s$ of the $x$-propagating A0 and S0 Lamb modes in an elastic aluminium plate $z \in (-d,d)$, see Fig.~\ref{fig3}(c). The parameters used here are: $f=20\,$kHz and $d=3\,$cm. 
(b) Schematics of the experimental setup for the spin-induced directional excitation of the Lamb modes. A circularly-polarized source with spin $s = +1$, placed at the edge of an aluminum plate, $x=0$, $z=-d$, excites oppositely-propagating A0 and S0 modes. 
(c) The experimentally measured Fourier $(\omega,k_x)$ spectra of the excited waves in the $x>0$ and $x<0$ zones of the plate versus the calculated spectra of the A0 and S0 Lamb modes. One can clearly see the spin-induced counter-directional propagation of the symmetric and antisymmetric modes.}
\label{fig4}
\end{figure}

{\it Anomalous spin-momentum locking in the thin-plate Lamb modes.---}
Importantly, the spin-sign anomalies in surface elastic waves are not merely curious facts but tools which could provide new functionalities to acoustic spin-based setups. To show this, we consider the spin-momentum locking using thin-plate modes. Such plates can support symmetric and anti-symmetric lowest-order modes: short-range and long-range surface plasmons for metalic films~\cite{Sarid1981} or S0 and A0 Lamb modes for elastic plates~\cite{Auld_book,Lanoy2020}. Akin to Figs.~\ref{fig1}(b)--(d), Figs.~\ref{fig3}(a)--(c) show the transverse spin densities $S$ for symmetric and antisymmetric electromagnetic, acoustic, and elastic modes. One can see that the electromagnetic and acoustic modes posses  similar spin distributions with no qualitative difference between the symmetric and anti-symmetric modes. In turn, the elastic case exhibits a remarkable anomaly: the spin distributions have {\it opposite signs} for the S0 and A0 Lamb modes. Figure~\ref{fig4}(a) shows that, akin to the Rayleigh-wave case, this anomaly is caused by the hybrid contributions $({s}_{lt} + {s}_{tl})$ to the spin density, which have an opposite sign and prevails over the pure contributions $({s}_{ll} + {s}_{tt})$ in the A0 mode, but not in the S0 mode.

Since the spin sign reverses with the reversal of the wave propagation direction, it determines
the intrinsic \textit{spin-momentum locking} in surface waves, with useful spin-to-direction coupling applications~\cite{Bliokh2015,Mechelen2016,Shi2019,Bliokh2019,Bliokh2015PR,Aiello2015,Rodriguez2013,Petersen2014,OConnor2014,Feber2015,Lodahl2017,Shi2019,Xu2020,Long2020,Yuan2021, Han2022}. This means that sources with opposite circular polarizations in the $(x, z)$ plane excite oppositely-propagating surface waves. Applying this to the modes in Fig.~\ref{fig3}, one can see that a circularly-polarized source located near one of the plate surfaces will excite symmetric and antisymmetric electromagnetic or acoustic modes propagating in the \textit{same} direction, but the S0 and A0 elastic modes propagating in \textit{opposite} directions. This provides a new tool for the efficient selective coupling to the symmetric and anti-symmetric Lamb waves.

We performed experiments confirming such selective excitation of the Lamb modes in an aluminum strip of thickness $2d = 6\,$cm. The experimental setup is shown in Fig.~\ref{fig4}(b). To create a circularly-polarized source, we installed a pair of PZT rings on two sides of a square aluminum block attached to the strip. The PZT rings can excite displacements perpendicular to its mounting surface, so that the two rings produced vibrations along the $x$ and $z$ axes. By adjusting the phase difference $\pm \pi /2$ between these vibrations, we obtained a right-hand/left-hand circularly polarized source with positive/negative spin ${s}$. 

In both cases this source excited elastic modes propagating in opposite $+x$ and $-x$ directions. Choosing a right-handed source with $s = +1$, we sent wavepacket-like signals with central frequency $f = 20\,$kHz and duration $T =2.5 \times 10^{-4}\,$s,
corresponding to a five-cycled tone burst pulse~\cite{Yuan2021}. 
By perforating small V-shape grooves in the sample \cite{Yuan2021} at $x=-45\,$cm and $x=45\,$cm we confirmed that the $+x$ and $-x$ propagating waves carry positive spin determined by the chiral source~\footnotemark[2].
To measure the A0/S0 character of the right-/left-propagating waves, we measured the displacement $u_z(t, x)$ on the surface of the plate (using the laser Doppler vibrometer) versus time and $x$-coordinate (with $1\,$cm intervals) and
Fourier-transformed the measured signals to the $(\omega, k_x)$ space. The results of such measurements in opposite $+x$ and $-x$ directions are shown in Fig.~\ref{fig4}(c). 
Comparison with the dispersion curves of the S0 and A0 Lamb modes~\footnotemark[2] clearly shows that our source with $s = +1$ excited the S0 mode propagating in the $-x$ direction and the A0 mode propagating in the $+x$ direction. This confirms the spin-controlled counter-directional excitation of the symmetric and anti-symemtric Lamb modes. 

The polarizations of the A0 and S0 modes at the plate edges are actually elliptical rather than perfectly circular. These polarizations depend on the frequency and can even reverse the direction of rotation, i.e., spin (for the S0 mode at $f\simeq 37\,$kHz). For details see the Supplementary Materials~\footnotemark[2]. In our experiment we chose the central frequency $f=20\,$kHz such that the edge polarizations are close enough to opposite circular polarizations, and that it is below the cutoff frequency of higher-order modes.

{\it Conclusions.---}
We have examined, both theoretically and experimentally, the unusual behavior of the transverse spin and spin-momentum locking in surface elastic waves. These properties exhibit several sign anomalies as compared with their electromagnetic, acoustic, and water-wave counterparts. We have shown that all these anomalies originate from the hybrid nature of surface elastic waves including the transverse (curl-less) and longitudinal (divergence-less) field contributions. We have experimentally measured the nontrivial distribution of the elastic spin density in surface Rayleigh waves. Furthermore, we have predicted and observed the unusual spin-momentum locking of the Lamb modes of a thin elastic plate. A chiral source located near an edge of the plate allows an efficient directional coupling to the counter-propagating symmetric S0 and antisymmetric A0 Lamb modes. Our results provide insights into fundamental physical properties of the spin and `hybrid spin' in elastic waves and offer new tools for spin-controlled manipulation of surface elastic waves in condensed-matter, metamaterial, and seismic systems.

\textcolor{black}{It is worth remarking that the normalized elastic spin, studied in this work, is closely related to the H/V ratio (ellipticity), which is one of the main parameters in studies of seismic Rayleigh waves \cite{Boore1969,Ferreira2007,Tanimoto2008,Lin2014,Berbellini2016}. In particular, the spin and H/V ratio have similar $z$-dependences and vanish in the same points. Our experimental technique can be further applied to inhomogeneous materials to compare with numerical calculations of the seismic H/V ratio and verification of the Earth models used there.} 


This work is supported by the National Key R\&D Program of China (2022YFA1404400), the National Natural Science Foundation of China (Nos. 11935010, and 12172256), the Natural Science Foundation of Shanghai (20ZR1462700, 23ZR1481200, 23XD1423800), and the Opening Project of Shanghai Key Laboratory of Special Artificial Microstructure Materials and Technology.
F.N. is supported in part by
Nippon Telegraph and Telephone Corporation (NTT) Research,
the Japan Science and Technology Agency (JST)
[via the Quantum Leap Flagship Program (Q-LEAP)], and
the Asian Office of Aerospace Research and Development (AOARD) (via Grant No. FA2386-20-1-4069).




%

\end{document}